\documentclass[english,aps,preprint,nofootinbib,superscriptaddress]{revtex4-2}%
\usepackage{mathpazo}
\usepackage[latin9]{inputenc}
\usepackage{geometry}
\usepackage{babel}
\usepackage{array}
\usepackage{verbatim}
\usepackage{float}
\usepackage{amsmath}
\usepackage{amssymb}
\usepackage{microtype}
\usepackage{amsfonts}
\usepackage{graphicx}
\usepackage{xcolor}%
\usepackage{chngcntr}

\begin{document}

\title{Characterizing entangled state update in different reference frames with weak measurements}
\author{J. Allam}

\affiliation{Laboratoire de Physique Th\'eorique et Mod\'elisation, CNRS Unit\'e 8089, CY
	Cergy Paris Universit\'e, 95302 Cergy-Pontoise cedex, France}
\author{ A. Matzkin}
\affiliation{Laboratoire de Physique Th\'eorique et Mod\'elisation, CNRS Unit\'e 8089, CY
	Cergy Paris Universit\'e, 95302 Cergy-Pontoise cedex, France}

\begin{abstract}
As per the projection postulate, the evolution of the quantum state of a system upon measurement results in state update. 
In this work, we investigate the characterization of updated states of multi-partite
entangled qubit states with non-destructive weak measurements, involving weakly
coupling a pointer to each qubit while considering a relativistic setup involving different reference frames. As is well-known, the updated state at intermediate times
is frame-dependent, and outcome randomness, intrinsic to projective measurements prevents
any information from being acquired on the updated state.
Here we will see that when weak measurements are implemented there is instead
an interplay between randomness and interference, depending
on the number of qubits, so as to
 maintain consistency between descriptions in arbitrary reference frames.\ Analytical
results are given for a small or infinite number of qubits, while
for a finite number of qubits we will resort to Monte-Carlo sampling in order
to simulate numerically the results of weak measurements. As a by-product, we
show how a single-shot measurement of the pointers' positions allows to make
quantitative predictions on the outcomes that can be obtained for any
observable measured on a distant qubit of the multi-partite state.

\end{abstract}
\maketitle

\newpage

\section{Introduction}

The tension between special relativity and quantum mechanics remains poorly
understood at a fundamental level \cite{gisin-science}. It is well-known from
Bell's theorem \cite{BT} that any rendering of the quantum correlations in
terms of underlying physical processes (``beables'' in Bell's terms) leads to a non-local model. 
In particular, if the quantum state is taken to be a ``beable'', then 
for a
multi-particle system in an entangled state,  we get the following features. First, 
a local measurement on one of the particles (say particle 1) instantaneously updates 
non-locally the state everywhere, including at space-like separated
points \cite{Albert-Aharonov-1,peres-intervention}. Second,
the updated state is only meaningful in a reference frame in which the measurement on
particle 1
happened first, given that according to special relativity, when different
measurements take place at space-like separated points, the time ordering of different events
can differ in distinct reference frames. While no-signaling ensures the
``peaceful coexistence'' \cite{shimony}, in
any reference frame, between special relativity and
quantum theory, Bell has argued \cite{bell-ns} that the impossibility of
sending signals can hardly be taken as the primary physical principle
accounting for the fundamental causal structure of the Universe.

Several works have examined the causality constraints on non-local
measurements \cite{Albert-Aharonov-1,Popescu-Vaidman,Reznik,brodutch}. In this
paper we will focus instead on these intermediate quantum states that appear,
in a given reference frame, as the result of state update subsequent to a
measurement on an entangled state. State update is a crucial ingredient in
several major quantum information protocols such as quantum teleportation, and
investigations concerning its implications have seen renewed interest
\cite{exp-time,srikanth,javurek,AMWF}. Consider for instance the bipartite entangled state
of two qubits A and B%
\begin{equation}
\left\vert \psi\right\rangle =\alpha\left\vert +u\right\rangle _{A}\left\vert
+w\right\rangle _{B}+\beta\left\vert -u\right\rangle _{A}\left\vert
+w\right\rangle _{B}+\gamma\left\vert +u\right\rangle _{A}\left\vert
-w\right\rangle _{B}\label{st1}%
\end{equation}
where $\left\vert \pm u\right\rangle $ and $\left\vert \pm w\right\rangle $
are the eigenstates of the spin projectors $\hat{\sigma}_{u}$ and $\hat
{\sigma}_{w}$ along two directions $u$ and $w$. Assume Alice and Bob are
space-like separated and measure respectively the qubits A and B in the
$\left\vert \pm u\right\rangle $ and $\left\vert \pm w\right\rangle $ basis.
In a reference frame in which Alice measures first, Bob should update
$\left\vert \psi\right\rangle $ to one of the two intermediate states
$\left\vert +w\right\rangle $ or $\left(  \alpha\left\vert +w\right\rangle
+\gamma\left\vert -w\right\rangle \right)  /\sqrt{\left\vert \alpha\right\vert
^{2}+\left\vert \gamma\right\vert ^{2}}$ depending on Alice's outcome. In a
reference frame in which Bob measures first, Alice should update $\left\vert
\psi\right\rangle $ to the intermediate states $\left\vert +u\right\rangle $
or $\left(  \alpha\left\vert +u\right\rangle +\beta\left\vert -u\right\rangle
\right)  \big/\sqrt{\left\vert \alpha\right\vert ^{2}+\left\vert
\beta\right\vert ^{2}}$. The intermediate states are different in each
reference frame, and they are not related by a Lorentz transform
\cite{peres-intervention}. Moreover, the consensus is that state update should
be viewed as taking place instantaneously in any reference frame
\cite{peres-review}. The reason is that other recipes (such as a 
light-cone dependent collapse \cite{HK}) were shown \cite{Albert-Aharonov-1} to fail, in
particular for measurements of non-local observables (but see Refs.
\cite{hiley,javurek} for more nuanced views). However, any attempt to characterize these quantum
states through standard projective measurements would fail, due to the
fundamental unpredictability intrinsic to quantum measurements.\ In this
sense, as it has often been argued, quantum randomness prevents signaling and
``saves'' relativistic causality \cite{PR,general}.

What happens if instead of standard projective measurements, we resort to
non-destructive measurements in order to attempt to characterize these
intermediate quantum states? What could in principle be observed in each
reference frame? How is non-signaling enforced in this case? In order to
tackle these questions, we will employ weak measurements \cite{AAV}, a
specific form of non-destructive measurements. The aim of a weak measurement
is to characterize the quantum state of a particle at an intermediate time,
between preparation and a final detection, by implementing a minimally
perturbing coupling between the particle and a quantum pointer. By
post-selecting the particle state, the pointer state acquires a shift that
depends on the weak value of the particle observable coupled to the pointer.
Although the precise physical meaning of weak values has remained
controversial \cite{bey,vaidman-con,FP,JJJ}, the operational implications of the weak measurement protocol are unambiguous: the
quantum state of the pointer is modified by a quantity that depends on the
quantum state of the particle at the time of the interaction.\ As such, weak
measurements have been used experimentally to characterize the
quantum states of a system at intermediate times, for example in order to
implement direct quantum state tomography or to observe the quantum state
evolution of a particle along different paths in an interferometer
\cite{lundeen,PRL127,ion,unambiguous}.

By conditioning the weak measurement process on a given post-selection,
randomness appears to be circumvented.\ However there is a trade-off: very
little information is acquired after a single measurement, so that a great
number of identical measurements must be made. This can simply involve
repeating the measurement so as to acquire sufficient statistics in order to
observe, in the weak measurement case, the shift of the quantum pointer.
Another option is to perform simultaneously a great number of measurements by
preparing identical pointers.\ In this work we will resort to both options.

Starting from a state of the type given by Eq. (\ref{st1}), we will suppose
that one of the observers, say Bob implements a standard projective
measurement on qubit B, while the other distant qubits are weakly
measured by Alice. In Sec. \ref{weakm}, we will consider a scenario in which 
Alice receives a single qubit. Even though Alice can repeat weak measurements
many times, no information can be acquired on the pre-selected state, that
varies randomly in one reference frame or is undefined in a reference frame in
which qubit B was not yet measured. In Sec. \ref{inf} we will examine the opposite
case, with Eq. (\ref{st1}) being replaced by a multi-partite entangled state so
that Alice can perform weak measurements on a huge number of weakly coupled
pointers so as to gather enough statistics in a single shot.
We will show however that although each qubit is subjected to a weak
measurement, the entire set of weakly measured qubits behaves collectively, as
the number of pointer tends to infinity, as a strongly coupled pointer,
limiting the information that can be obtained from the measurement.\ Finally
in Sec. \ref{sev} we will consider a smaller (``finite'') number of pointers and model numerically the output
of the weakly coupled pointer measurements by resorting to numerical simulations 
through Monte-Carlo
sampling. In this regime, statistics obtained from a single shot measurement on the pointers allow Alice to make 
quantitative predictions on qubit B's outcome for an arbitrary observable.  We will discuss our results and conclude in Sec. \ref{disc}, stressing in particular that only correlations, valid in any reference frame and relevant to the full quantum state, can be characterized with weak measurements.

\section{Weak measurement of a single qubit on multiple copies}

\label{weakm}

Let us consider two qubits A and B initially ($t=0$) prepared in state
$\left\vert \psi\right\rangle $ given by Eq. (\ref{st1}), each particle flying
in opposite directions.\ At $t=t_{B}$ Bob (located at $x_{B})$ receives his
qubit and measures $\hat{\sigma}_{w}$. For $t>t_{B}$ the state of particle A
is updated to either
\begin{equation}
\left\vert \psi_{A}^{+}\right\rangle =\left\vert +u\right\rangle \label{psp}%
\end{equation}
or%
\begin{equation}
\text{\ }\left\vert \psi_{A}^{-}\right\rangle =\frac{\alpha\left\vert
+u\right\rangle +\beta\left\vert -u\right\rangle }{\sqrt{\left\vert
\alpha\right\vert ^{2}+\left\vert \beta\right\vert ^{2}}}. \label{psm}%
\end{equation}
At $t_{A}>t_{B}$ Alice, located at $x_{A}$ performs a weak measurement of some
observable $\hat{O}$ by coupling qubit A to a weak pointer initially prepared
in state $\left\vert \varphi\right\rangle $, typically a Gaussian function
centered at $X=0$. The first step of the weak measurement, resulting from the
qubit-pointer interaction Hamiltonian, is described by the unitary
$\exp(-ig\hat{O}\hat{P})\left\vert \psi_{A}^{\pm}\right\rangle \left\vert
\varphi\right\rangle $ where $\hat{P}$ is the momentum of the pointer and $g$
the coupling strength.\ As required by the weak measurements formalism, $g$
must be very small, so that asymptotic expansions of the interaction unitary
hold. After the weak interaction, the qubit is immediately post-selected to
the eigenstate $\left\vert f\right\rangle $ of another qubit
observable.\ Hence after post-selection the (unnormalized) state of the pointer
is either \cite{AAV,FP}%
\begin{equation}
\left\vert \varphi_{A}^{+}\right\rangle =\exp\left(  -i\epsilon_{+}\right)
\left\vert \varphi\right\rangle \equiv\left\vert \varphi^{\epsilon_{+}%
}\right\rangle
\end{equation}
or%
\begin{equation}
\left\vert \varphi_{A}^{-}\right\rangle =\left(  \alpha\exp\left(
-i\epsilon_{+}\right)  +\beta\exp\left(  -i\epsilon_{-}\right)  \right)
\left\vert \varphi\right\rangle \equiv\alpha\left\vert \varphi^{\epsilon_{+}%
}\right\rangle +\beta\left\vert \varphi^{\epsilon_{-}}\right\rangle ,
\label{phiAmin}
\end{equation}
where we have denoted the shifts by $\epsilon_{\pm}$,%
\begin{equation}
\epsilon_{\pm}=g\operatorname{Re}O_{\pm}^{w} \label{epsdef}%
\end{equation}
with $O_{\pm}^{w}$ being the weak value%
\begin{equation}
O_{\pm}^{w}=\frac{\left\langle f\right\vert \hat{O}\left\vert \pm
u\right\rangle }{\left\langle f\right\vert \left.  \pm u\right\rangle }.
\label{wv0}%
\end{equation}

For simplicity, we will assume in this work that the weak values are real. Note that $\left| \psi_A^- \right\rangle$ can also be expressed in terms of a single shifted pointer state, with the weak value \cite{DMMJB, TC}
\begin{equation}
   \frac{  \left\langle f \left| \hat{O} \right| \psi_{A}^- \right\rangle }{  \left\langle  f | \psi_{A}^- \right\rangle}.
\end{equation} 
 It will be convenient for what follows to use the form given by the right hand side of Eq. (\ref{phiAmin}).

The two pointer states $\varphi_{A}^{+}(X)$ and $\varphi_{A}^{-}(X)$ are
slightly different -- they nearly overlap but present different shifts. In
order to determine the profiles experimentally, several position measurements
of the pointer must be made, hence the need of multiple copies. However this
procedure succeeds only if the initial (\textquotedblleft
pre-selected\textquotedblright) state, $\left\vert \psi_{A}^{+}\right\rangle $
or $\left\vert \psi_{A}^{-}\right\rangle $ is known. This is possible but requires classical communication
from Bob.\ Otherwise the updated state remains unknown and
Alice deals with a statistical mixture of pointer states from which no
information on Bob's measurement can be retrieved.

\begin{figure}[H]
\centering
\includegraphics[width=0.6\textwidth]{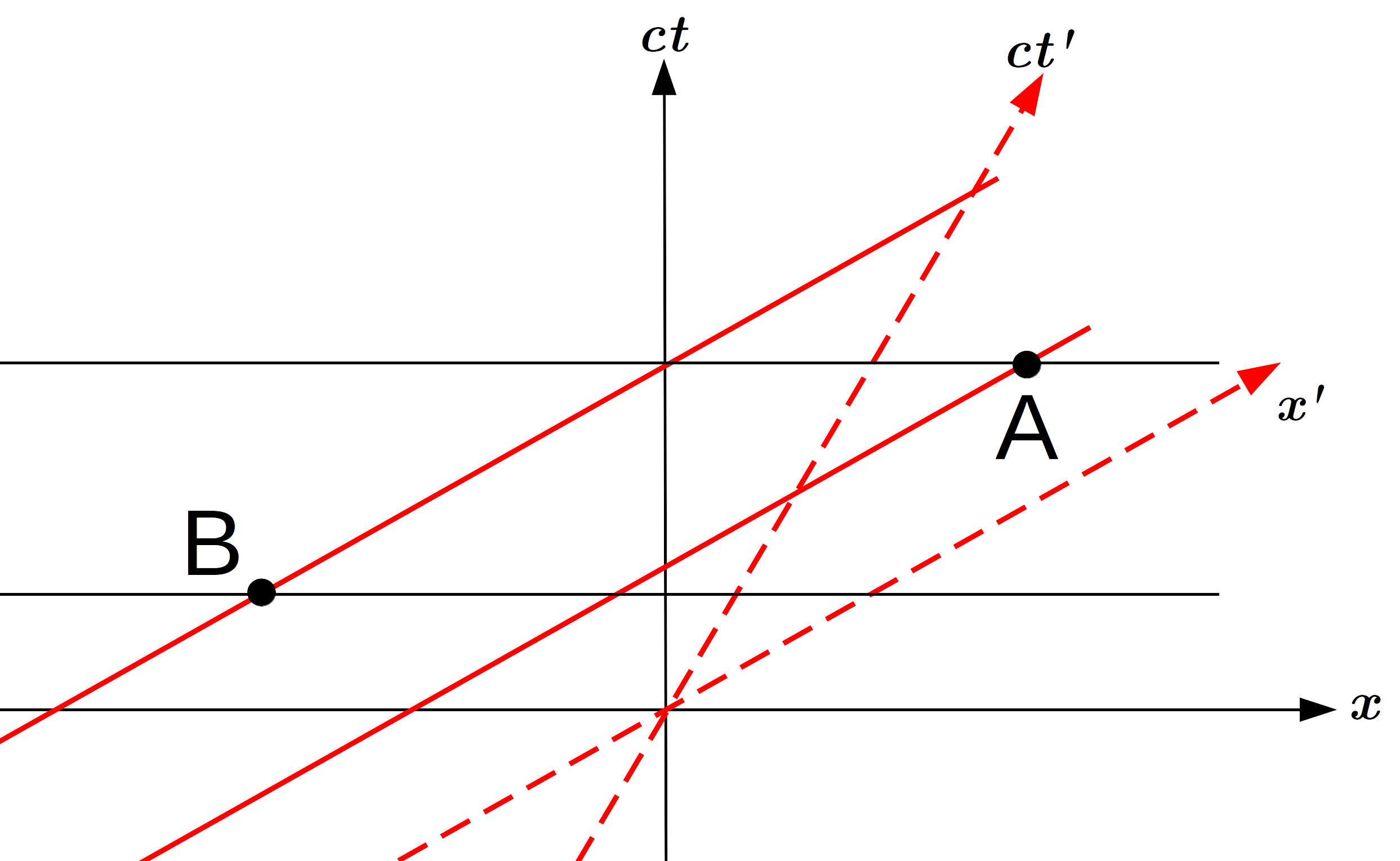}\caption{Instantaneous state
update in two reference frames. In frame $\mathcal{R}$ (black) Bob measures
first updating the state to $\left\vert \psi_{A}^{\pm}\right\rangle $, while
in frame $\mathcal{R^{\prime}}$ (red) Alice measures first updating the state
to $\big\vert \psi_{B}^{X_{m}}\big\rangle $.}%
\label{fig-spacetime}%
\end{figure}

If $(t_{B},x_{B})$ and $\left(  t_{A},x_{A}\right)  $ are space-like
separated, there is a reference frame $\mathcal{R}^{\prime}$ in which the time
ordering of the measurements is inverted, $t_{A}^{\prime}<t_{B}^{\prime}$
(see Fig. \ref{fig-spacetime}). In $\mathcal{R}^{\prime},$ the qubit A is not
in a pre-selected state, given that the initial state is the entangled state
$\left\vert \psi\right\rangle .$ An observer Alice$^{\prime}$ in
$\mathcal{R}^{\prime}$ can of course trace out particle B and obtain
statistics for particle A measurements, but in terms of an actual outcome
the only fact is the measured spatial position $X_{m}$ of the pointer.\ After the
pointer's measurement, qubit B is updated to the state in $\mathcal{R}^{\prime}$ (here left unnormalized)  %
\begin{equation}
\left\vert \psi_{B}^{X_{m}}\right\rangle =\alpha\left\langle X_{m}
\vert\varphi^{\epsilon_{+}}\right\rangle \left\vert +w\right\rangle _{B}%
+\beta\left\langle X_{m} \vert\varphi^{\epsilon_{-}}\right\rangle \left\vert
+w\right\rangle _{B}+\gamma\left\langle X_{m} \vert\varphi^{\epsilon_{+}%
}\right\rangle \left\vert -w\right\rangle _{B} \label{update-1wm}%
\end{equation}

$\left\vert \psi_{B}^{X_{m}}\right\rangle $ can be any state compatible with
the distribution of the continuous random variable $X_{m}$. At $t_{B}^{\prime
}$ an observer Bob$^{\prime}$ sees qubit B being measured in the $\pm w$
basis.\ If the procedure is repeated, Bob$^{\prime}$ would infer that qubit B
is updated each time to a different random quantum state. Classical
communication from Bob$^{\prime}$ to Alice$^{\prime}$ is received in
$\mathcal{R}^{\prime}$ after both qubits have been measured. With this
communication Alice$^{\prime}$ can sort her past results and recover
statistically the correlations embodied in the entangled state $\left\vert
\psi\right\rangle $. But no information can be extracted on the intermediate state.

We see that it is not possible to characterize the intermediate updated state
in $\mathcal{R}$ by weakly measuring a single qubit on multiple copies -- as
usual no information can be extracted after a single weak measurement, while
repeating the procedure on multiple copies conflicts with the random outcomes
of qubit B's measurements. The interplay of the minimally perturbing character
of weak measurements and the randomness of projective measurements makes the
description compatible in both reference frames.

\section{Many weakly coupled pointers: the infinite limit}

\label{inf}

\subsection{Multipartite states}

Consider a multipartite entangled state similar to Eq. (\ref{st1}) but with a
huge number of qubits in the same state sent to Alice's side,%
\begin{equation}
\left\vert \psi\right\rangle =\alpha\bigotimes_{i=1}^{N}\left\vert
+u\right\rangle _{i}\left\vert +w\right\rangle _{B}+\beta\bigotimes_{i=1}%
^{N}\left\vert -u\right\rangle _{i}\left\vert +w\right\rangle _{B}%
+\gamma\bigotimes_{i=1}^{N}\left\vert +u\right\rangle _{i}\left\vert
-w\right\rangle _{B}. \label{many}%
\end{equation}
One weak measurement does not give enough information and must be
repeated, but $N$ identical weak measurements (with $N\rightarrow\infty$)
do.\ Indeed, assume $\alpha=0$ in Eq. (\ref{many}); qubit B is measured first
so that after state update Alice receives either $\bigotimes_{i=1}%
^{N}\left\vert +u\right\rangle _{i}$ or $\bigotimes_{i=1}^{N}\left\vert
-u\right\rangle _{i}$.\ Assume further that each qubit in state $\left\vert
\pm u\right\rangle _{i}$ is weakly coupled to a quantum pointer in state
$\left\vert \varphi_{i}\right\rangle .\ $Let us choose the post-selected
state, identical for each qubit, such that $\left\vert \left\langle
f\right\vert \left.  +u\right\rangle \right\vert =\left\vert \left\langle
f\right\vert \left.  -u\right\rangle \right\vert $.\ Then after post-selection
each pointer is slightly shifted, $\left\vert \varphi_{i}\right\rangle
\rightarrow\left\vert \varphi_{i}^{\epsilon_{\pm}}\right\rangle $ \ Using the
notation
\begin{equation}
\left\vert \Phi^{\epsilon_{\pm}}\right\rangle \equiv%
{\displaystyle\prod_{i=1}^{N}}
\left\vert \varphi_{i}^{\epsilon_{\pm}}\right\rangle , \label{def-comp}%
\end{equation}
we end up with a wavefunction for the shifted pointers that is either
$\Phi^{\epsilon_{+}}(x_{1},...,x_{N})$ or $\Phi^{\epsilon-}(x_{1},...,x_{N}%
)$.\ By measuring the position of each identical pointer, Alice has enough
statistics to reconstruct the profile and determine the shift $\epsilon_{+}$
or $\epsilon_{-}$, thereby discriminating $\bigotimes_{i=1}^{N}\left\vert
+u\right\rangle _{i}$ from $\bigotimes_{i=1}^{N}\left\vert -u\right\rangle
_{i}$ in a single experiment (note that the two qubit states to be
discriminated need not be orthogonal). Notice that this is not particularly
surprising, after all the same conclusion can be reached by projective
measurements of the $N$ qubits.

Now if $\alpha\neq0$, the updated states become
\begin{equation}
\left\vert \psi_{A}^{+}\right\rangle =\frac{1}{\sqrt{\left\vert \alpha
\right\vert ^{2}+\left\vert \beta\right\vert ^{2}}}\left(  \alpha
\bigotimes_{i=1}^{N}\left\vert +u\right\rangle _{i}+\beta\bigotimes_{i=1}%
^{N}\left\vert -u\right\rangle _{i}\right)  \label{superi}%
\end{equation}
and%
\begin{equation}
\left\vert \psi_{A}^{-}\right\rangle =\bigotimes_{i=1}^{N}\left\vert
+u\right\rangle _{i}.
\end{equation}
Introducing the $N$ weakly coupled pointers leads after post-selection to%
\begin{align}
\left\vert \Phi_{A}^{+}\right\rangle  &  =\alpha\left\vert \Phi^{\epsilon_{+}%
}\right\rangle +\beta\left\vert \Phi^{\epsilon-}\right\rangle \label{superif}%
\\
\left\vert \Phi_{A}^{-}\right\rangle  &  =\left\vert \Phi^{\epsilon_{+}%
}\right\rangle , \label{superif2}%
\end{align}
where for simplicity we have assumed $\left\langle f\right\vert \left.
+u\right\rangle =\left\langle f\right\vert \left.  -u\right\rangle $ and left
as usual the pointer states unnormalized. Now discriminating $\left\vert
\Phi_{A}^{+}\right\rangle $ from $\left\vert \Phi_{A}^{-}\right\rangle $ (or
more generally from another linear superposition $\gamma\left\vert
\Phi^{\epsilon_{+}}\right\rangle +\delta\left\vert \Phi^{\epsilon
-}\right\rangle $) would reveal instantaneously qubit B's measurement
basis and outcome.\ In this situation, no-signaling is enforced by a rather
curious property.
 
\subsection{From nearly overlapping to quasi-orthogonal pointers}

Recall that an individual pointer state is minimally shifted after a weak
measurement, so that two quantum states $\left\vert \varphi^{\epsilon
}\right\rangle $ and $\left\vert \varphi^{\epsilon^{\prime}}\right\rangle $
are almost perfectly overlapping
\begin{equation}
\big\langle\varphi^{\epsilon}|\varphi^{\epsilon^{\prime}}\big\rangle\approx
1-i\left(  \epsilon^{\prime}-\epsilon\right)  \big\langle\varphi
\big\vert P\big\vert\varphi\big\rangle, \label{overlap}%
\end{equation}
where $\epsilon$ and $\epsilon^{\prime}$ involve weak values such as $O_{\pm
}^{w}$ that might differ only in the pre-selected state [see Eqs.
(\ref{epsdef}) and (\ref{wv0})]. However the tensor product of $N$ identically
shifted pointers becomes orthogonal, in the limit of large $N$ to another
product of shifted states,%
\begin{equation}
\big\langle\Phi^{\epsilon}\big|\Phi^{\epsilon^{\prime}}\big\rangle\underset
{N\rightarrow\infty}{\rightarrow}0. \label{vanish}%
\end{equation}
where we follow the notation introduced in Eq. (\ref{def-comp}). This can easily be seen if $\varphi^{\epsilon}(x)$ is taken to be a Gaussian,
since $\big\langle\varphi^{\epsilon}|\varphi^{\epsilon^{\prime}}%
\big\rangle=\exp\left(  -\left(  \epsilon-\epsilon^{\prime}\right)  ^{2}%
/8d^{2}\right)  \approx1$ ($d$ is the Gaussian width which must be large
relative to the shifts in the weak regime \cite{sudarshan}), but for $N$
pointers, $\big\langle\Phi^{\epsilon}|\Phi^{\epsilon^{\prime}}\big\rangle=\exp
\left(  -N\left(  \epsilon-\epsilon^{\prime}\right)  ^{2}/8d^{2}\right)
\rightarrow0$ for sufficiently large $N$.

Actually, one can prove that in the $N\rightarrow\infty$ limit the shift
$\epsilon$ becomes an eigenvalue of the observable $\hat{\xi}=\sum_{i}\hat
{X}_{i}/N,$%
\begin{equation}
\hat{\xi}\left\vert \Phi^{\epsilon}\right\rangle =\epsilon\left\vert
\Phi^{\epsilon}\right\rangle \label{ortho}%
\end{equation}
where $\hat{X}_{i}$ is the position of the $i$th pointer. This can be seen by
using the relation (see e.g. \cite{tollaks} for the proof)
\begin{equation}
\hat{X}_{i}\left\vert \varphi_{i}\right\rangle =\left\langle \hat{X}%
_{i}\right\rangle \left\vert \varphi_{i}\right\rangle +\Delta\hat{X}%
_{i}\left\vert \varphi_{i}^{\perp}\right\rangle \label{rel}%
\end{equation}
where $\left\vert \varphi_{i}^{\perp}\right\rangle $ is defined from
$\big\langle\varphi_{i}^{\perp}\big\vert\varphi_{i}\big\rangle=0$ and is not
normalized; here $\left\langle \hat{X}_{i}\right\rangle =\left\langle
\varphi_{i}\right\vert \hat{X}_{i}\left\vert \varphi_{i}\right\rangle $,
$\Delta\hat{X}_{i}=\left\langle \varphi_{i}\right\vert \left\langle \hat
{X}_{i}^{2}\right\rangle -\left\langle \hat{X}_{i}\right\rangle ^{2}\left\vert
\varphi_{i}\right\rangle ^{1/2}$. Using the notation $\left\vert \Phi
_{i}^{\epsilon}\right\rangle ^{\perp}=\left\vert \varphi_{1}^{\epsilon
}\right\rangle \otimes\left\vert \varphi_{2}^{\epsilon}\right\rangle
\otimes...\otimes\left\vert \varphi_{i}^{\epsilon\perp}\right\rangle
\otimes...\otimes\left\vert \varphi_{N}^{\epsilon}\right\rangle $ we apply Eq.
(\ref{rel}) to $\left\vert \Phi^{\epsilon}\right\rangle $ with
\begin{equation}
\left\vert \Phi^{\epsilon}\right\rangle ^{\perp}=\sum_{i}\left\vert \Phi
_{i}^{\epsilon}\right\rangle ^{\perp}%
\end{equation}
yielding
\begin{equation}
\hat{\xi}\big\vert\Phi^{\epsilon}\big\rangle=\left\langle \hat{X}\right\rangle
\big\vert\Phi^{\epsilon}\big\rangle+\frac{\Delta\hat{\xi}}{\sqrt{N}}\sum
_{i=1}^{N}\big\vert\Phi_{i}^{\epsilon}\big\rangle^{\perp}. \label{xi}%
\end{equation}
The first term on the right hand side is independent of $N$ and is obtained by
remarking that $\left\langle \hat{\xi}\right\rangle =\sum_{i=1}%
^{N}\left\langle \hat{X}_{i}\right\rangle /N=\left\langle \hat{X}\right\rangle
$.\ In the second term, $\Delta\hat{\xi}$ is also independent of $N$, but the
$1/\sqrt{N}$ factor, that comes about by normalizing $\left\vert
\Phi^{\epsilon}\right\rangle ^{\perp}$ to 1 (since $\left\langle
\Phi^{\epsilon}\right\vert \left.  \Phi^{\epsilon}\right\rangle ^{\perp
}=\left\langle \Phi^\epsilon\right\vert \hat{\xi}\left\vert \Phi^{\epsilon
}\right\rangle ^{\perp}$) renders the second term negligible as $N\rightarrow
\infty$. Equations similar to Eq. (\ref{xi}) were employed previously in the context of 
investigations of the macroscopic limit of a huge number of identical systems
\cite{ahar-reznik,finkelstein}.

The upshot is that when the number of pointers is large, the states
$\left\vert \Phi^{\epsilon}\right\rangle $ and $\left\vert \Phi^{\epsilon
^{\prime}}\right\rangle $ become quasi-orthogonal despite the fact that the
shifts $\epsilon$ and $\epsilon^{\prime}$ are asymptotically small. Therefore
superposition of pointers such as Eq. (\ref{superif}) cannot be discriminated
in one shot -- actually measuring weakly a very large number of pointers is
equivalent to a strong measurement of a single qubit.\ Measuring the pointers
average in a state $\alpha\left\vert \Phi^{\epsilon_{+}}\right\rangle
+\beta\left\vert \Phi^{\epsilon-}\right\rangle $ invariably yields
$\epsilon_{+}$ or $\epsilon_{-}$. We see that randomness enters again so as to
prevent any information to be extracted from the intermediate state, though
randomness appears now in a peculiar way: rather than setting the outcome of a
single qubit, the random measured positions of each pointer ``conspire'' to select one of the compound states $\left\vert
\Phi^{\epsilon_{\pm}}\right\rangle $ of the superposition.

This ``conspiracy'' is necessary in order to
maintain consistency between observers in different reference frames. In the
frame $\mathcal{R}^{\prime}$ introduced in Sec. \ref{weakm}, the intermediate
states $\left\vert \psi_{A}^{\pm}\right\rangle $ or $\left\vert \Phi_{A}^{\pm
}\right\rangle $ of Eqs. (\ref{superi})-(\ref{superif2}) do not exist. In
$\mathcal{R}^{\prime}$ an observer Alice$^{\prime}$ would see the $N\ $qubits
measured before qubit B's measurement takes place. After coupling and
post-selection, the initial state $\left\vert \Psi(t=0)\right\rangle
=\left\vert \psi\right\rangle \left\vert \Phi\right\rangle $ (where
$\left\vert \psi\right\rangle $ is given by Eq. (\ref{many}) and $\left\vert
\Phi\right\rangle =\left\vert \varphi_{1}\right\rangle ...\left\vert
\varphi_{N}\right\rangle $ is the ready state of the $N$ pointers) becomes%
\begin{equation}
\left\vert \Psi(t_{1}^{\prime})\right\rangle =\left\vert \Phi^{\epsilon_{+}%
}\right\rangle \left(  \alpha\left\vert +w\right\rangle _{B}+\gamma\left\vert
-w\right\rangle _{B}\right)  +\beta\left\vert \Phi^{\epsilon-}\right\rangle
\left(  \left\vert +w\right\rangle _{B}\right)  . \label{bigpsi1}%
\end{equation}
After the pointers are measured, due to the quasi-orthogonality condition
(\ref{ortho}), qubit B's state is updated to one of the states between $(..)$
in Eq. (\ref{bigpsi1}), similarly to a standard projective measurement. This is different
than the update to state $\left\vert \psi_{B}^{x_{m}}\right\rangle $ of Eq.
(\ref{update-1wm}) when a single qubit was weakly measured -- the pointer was
then found to be in position $X_{m}$ and qubit B's state updated according to a state depending on $X_m$.

\section{A few weakly coupled pointers: Monte-Carlo simulations}

\label{sev}


\subsection{Setting}

We will now seek to investigate in this section the intermediate regime
between the single weakly coupled pointer of Sec. \ref{weakm} and the large
(\textquotedblleft infinite\textquotedblright) number of pointers of Sec.
\ref{inf}. In this case, when $N$ is of the order of a few tens or a few
hundreds, the resulting states $\left\vert \Phi^{\epsilon}\right\rangle $ are
not orthogonal, so there should be signatures of the
superpositions of such states. The drawback is that in this regime it is not
obvious to obtain analytical results for a one shot
measurement of $N$ weakly coupled pointers. We will need to resort to
numerical experiments, which we will do in the form of Monte-Carlo
simulations. More precisely, given a generic state of the form $\alpha
\left\vert \Phi^{\epsilon_{+}}\right\rangle +\beta\left\vert \Phi^{\epsilon
-}\right\rangle $ (where $\left\vert \Phi^{\epsilon}\right\rangle $ is defined
as per Eq. (\ref{def-comp}) above), the Monte-Carlo simulations will be
employed to generate sets of positions $X_{1},...,X_{N}$ representing the
outcomes of the pointers' position measurements.

\subsection{Pointer states in $\mathcal{R}$}

The scenario remains the same as in Sec. \ref{inf}, with the qubits prepared
in the entangled state $\left\vert \psi\right\rangle $ given by Eq.
(\ref{many}). In $\mathcal{R}$ qubit B is measured first and Alice applies
state update and couples $N$ pointers to the $N$ qubits she received. To be
specific, let us assume Bob has two choices of spin measurement, $\hat{\sigma
}_{w}$ or $\hat{\sigma}_{\theta}$ with the corresponding eigenstates denoted
$\left\vert \pm w\right\rangle $ and $\left\vert \pm\theta\right\rangle $. Let
us also suppose each pointer to be initially a Gaussian wavefunction $\langle
X_{i}|\varphi_{i}\rangle=\exp\left(  -X_{i}^{2}/4d^{2}\right)  /(2\pi
d^{2})^{1/4}$ (from now on, we will measure position in units of $d$, i.e., we
set $d=1)$. After the weak measurement of each of the $N\ $qubits$\ $the
updated state is one of the following
\begin{align}
\left\vert \Phi_{A}^{+w}\right\rangle  &  =\alpha\big\vert\Phi^{\epsilon_{+}%
}\big\rangle+\beta\big\vert\Phi^{\epsilon_{-}}\big\rangle\label{fipw}\\
\left\vert \Phi_{A}^{-w}\right\rangle  &  =\big\vert\Phi^{\epsilon_{+}%
}\big\rangle\label{fimw}\\
\left\vert \Phi_{A}^{+\theta}\right\rangle  &  =\bigg(\alpha\langle
+\theta|+w\rangle+\gamma\langle+\theta|-w\rangle\bigg)\big\vert\Phi
^{\epsilon_{+}}\big\rangle+\beta\langle+\theta|+w\rangle\big\vert\Phi
^{\epsilon_{-}}\big\rangle\label{fipt}\\
\left\vert \Phi_{A}^{-\theta}\right\rangle  &  =\bigg(\alpha\langle
-\theta|+w\rangle+\gamma\langle-\theta|-w\rangle\bigg)\big\vert\Phi
^{\epsilon_{+}}\big\rangle+\beta\langle-\theta|+w\rangle\big\vert\Phi
^{\epsilon_{-}}\big\rangle, \label{fimt}%
\end{align}
where we used the same notation as in Eqs. (\ref{superif})-(\ref{superif2}).
Alice measures the position of each pointer, acquiring in one shot the
positions $\{X_{1},...,X_{N}\}$.

\subsection{Monte-Carlo sampling}

The one-shot pointer positions, which form a statistical ensemble of continuous random variables $\mathbb{X}=\{X_{1},...,X_{N}\}$ are generated from
Monte-Carlo simulations, described in the Appendix.\ The target function (which is the probability density function we wish to visualize by the Monte-Carlo sampling) is one of the
four distributions $\left\vert \left\langle X_{1},...,X_{N}\right\vert \left.
\Phi_{A}^{\pm w}\right\rangle \right\vert ^{2}$ or $\left\vert \left\langle
X_{1},...,X_{N}\right\vert \left.  \Phi_{A}^{\pm\theta}\right\rangle
\right\vert ^{2}$ arising from the pointer wavefunctions (\ref{fipw}%
)-(\ref{fimt}).\ For a given target function, we generate $k$ sets
$\mathbb{X}^{k}$ each corresponding to the realization of a one-shot
measurement of the $N$\ pointers (an example of a realization is given in Fig. (\ref{sample-fig})). For a given realization $k$ an observer such
as Alice can compute simple statistics such as the average pointer position $\xi^{k}=\sum
_{i=1}^{N}X_{i}^{k}/N$ and the root-mean square. Averaging over $k$, one
expects to recover the quantum mechanical averages, e.g. $\langle\xi
\rangle=\langle\Phi|\xi|\Phi\rangle$ for the average of the mean position;
since we have assumed Gaussian pointers, it is straightforward to obtain
$\langle\xi\rangle$ or the variance $\left(  \Delta\xi\right)  ^{2}%
=\langle\Phi|\xi^{2}|\Phi\rangle-\langle\Phi|\xi|\Phi\rangle^{2}$ analytically
for a wavefunction of the form $\big\vert\Phi\big\rangle=a\big\vert\Phi
^{\epsilon_{+}}\big\rangle+b\big\vert\Phi^{\epsilon_{-}}\big\rangle.$

We will present numerical results for the choice $\alpha=\gamma=1/\sqrt{12}$
\ and $\beta=\sqrt{5/6}$; for the measurement basis we set as $\hat{\sigma
}_{w}=\hat{\sigma}_{z}$ and $\hat{\sigma}_{\theta}=\hat{\sigma}_{\pi/4}$ for
Bob and $\hat{O}=\hat{\sigma}_{z}$ and $\left\vert f\right\rangle =\left\vert
+x\right\rangle $ for Alice's weakly measured observable and post-selection
state respectively (we denote $|\pm x\rangle=\frac{1}{\sqrt{2}}(|+\rangle
\pm|-\rangle)$, where ${|\pm\rangle}$ constitute the eigenbasis of $\sigma_z$). This choice arises from a careful tuning in order to get a
representative situation. The initial state (\ref{many}) then takes the form
\begin{equation}
\left\vert \psi\right\rangle =\frac{1}{\sqrt{6}}\bigotimes_{i=1}^{N}\left\vert
+\right\rangle _{i}\left\vert +x\right\rangle _{B}+\sqrt{\frac{5}{6}%
}\bigotimes_{i=1}^{N}\left\vert -\right\rangle _{i}\left\vert +\right\rangle
_{B}, \label{manysp}%
\end{equation}
the weak values (\ref{wv0}) are readily computed as $\langle+x|\sigma_{z}%
|\pm\rangle/\langle+x|\pm\rangle=\pm1$ and the shifts (\ref{epsdef}) become
$\epsilon^{\pm}=\pm g$. A Monte-Carlo simulation can then be undertaken for
each of the four updated states (\ref{fipw})-(\ref{fimt}) . We present here
results for different values of the number of qubits $N$.

\begin{figure}[H]
\centering
\includegraphics[width=0.49\textwidth]{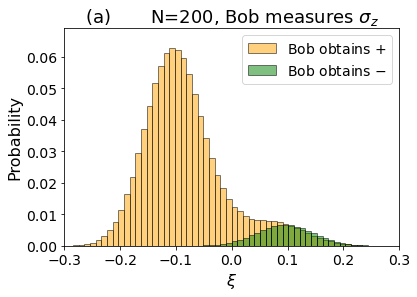}
\hfill\includegraphics[width=0.49\textwidth]{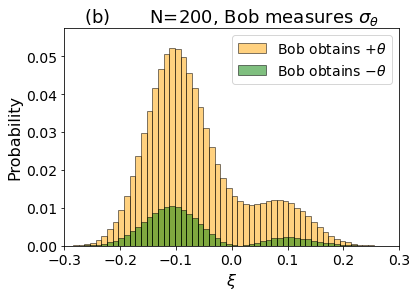}
\caption{Probability distribution of $\xi$ obtained from Monte-Carlo
simulations for $N=200$ weak pointers after state update in the frame
$\mathcal{R}$ (where qubit B is measured first). Bob chooses one of two
measurements; either along the z direction or along a direction $\theta=\pi
/4$. (a) Qubit B has been measured in the $z$-basis. If the outcome is
$^{\prime}+^{\prime}$, $\xi$ obeys the orange colored distribution, and if the
outcomes is $^{\prime}-^{\prime}$ it obeys the green colored distribution. (b)
Same as (a) when qubit B has been measured in the $\theta$-basis. The weak
coupling constant is set to $g=0.1$ ($g$ is given in units of position, which is in turn given in units of the Gaussian width $d$ (we set $d=1$)).}%
\label{fig-N200}%
\end{figure}

Fig. \ref{fig-N200} shows the distribution of $\xi_{k}$ for $N=200$ pointers
when the updated state is $\left\vert \Phi_{A}^{\pm z}\right\rangle $ (Fig.
\ref{fig-N200}(a)) or $\left\vert \Phi_{A}^{\pm\pi/4}\right\rangle $ (Fig.
\ref{fig-N200}(b)). The probability densities shown in the plots take into
account the probability to obtain a given state update (i.e., a given qubit B
measurement outcome). Notice that if Bob's measurement is known to be
$\hat{\sigma}_{z}$, then Fig. \ref{fig-N200}(a) tells us that there is a range
of $\xi$ values, $\xi\lesssim-0.1$, for which a single shot measurement of the
pointers gives unambiguous information on Bob's outcome simply by computing
the average of the $N$ pointers' position. This is similar to the $N\rightarrow
\infty$ limit of Sec.\ \ref{inf}, or for that matter to a projective
measurement of $\hat{\sigma}_{u}$ on the bipartite entangled state of Eq.
(\ref{st1}). For values $\xi\gg-0.1$ the distributions corresponding to the
states $\left\vert \Phi_{A}^{\pm z}\right\rangle $ overlap and a single shot
measurement of the $N$ pointers does not discriminate between both
possibilities, similarly to the overlap between slightly shifted pointers in
the case of a single weak measurement (Sec. \ref{weakm}). However for each
observed value of $\xi$ it is possible to quantify the relative probabilities
of qubit B's past or future outcome: for instance if $\xi\gtrsim0.1$ is
observed, one infers that the outcomes $\pm1$ are roughly equiprobable.

Fig.\ \ref{fig-N200}(b) shows similar features, except that there is a
substantial overlap between the measured updated states $\left\vert \Phi
_{A}^{\pm\pi/4}\right\rangle $ and that the $+1$ outcome is much more
probable; note that when the pointer average is $\xi\simeq0.04$ can it be said
with nearly unit probability that the updated state was $\left\vert \Phi_{A}^{+\pi
/4}\right\rangle $ given that the probability density for $\left\vert \Phi
_{A}^{-\pi/4}\right\rangle $ is nearly zero around $\xi\simeq0.04$.

\begin{figure}[H]
\centering
\includegraphics[width=0.49\textwidth]{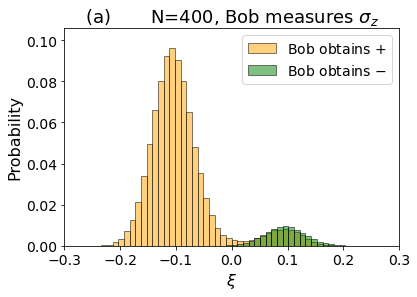}
\hfill\includegraphics[width=0.49\textwidth]{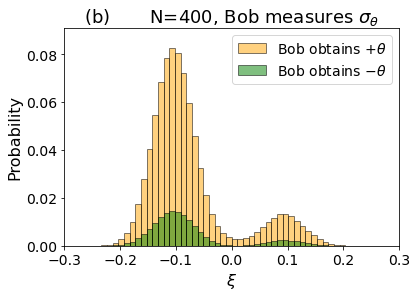} \caption{Same as
Fig. \ref{fig-N200} but for $N=400$ pointers. }%
\label{400}%
\end{figure}

If we increase the number of qubits that Alice measures, the overlap is
reduced until we get a behavior approaching the $N\rightarrow\infty$ limit. In
Fig. \ref{400}, where $N=400$, we see that the overlap between the peaks
corresponding to $\left\vert \Phi_{A}^{\pm z}\right\rangle $ or $\left\vert
\Phi_{A}^{\pm\pi/4}\right\rangle $ is starting to decrease, but the regime is
still similar to the $N=200$ case: If Bob measures in the $\pm$ basis, Alice
can be certain of his outcome only if she obtains $\xi\lesssim0$, and if Bob
chooses to measure in the $\pm\theta$ basis, Alice can be certain of his
outcome only if she obtains an average position over her pointers of
$\xi\simeq0.01$. 


\subsection{Pointer states in $\mathcal{R}^{\prime}$}

\begin{figure}[H]
\centering
\includegraphics[width=0.6\textwidth]{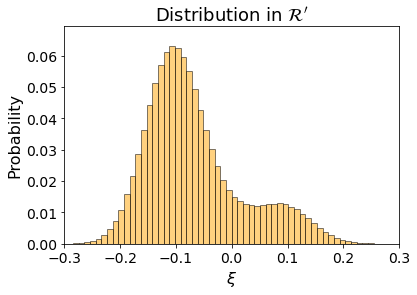}\caption{Probability
distribution of $\xi$ for $N=200$ in the reference frame $\mathcal{R^{\prime}}$.}%
\label{fig-Rprim}%
\end{figure}

In $\mathcal{R}^{\prime}$ the pointers are measured in the entangled state
$\left\vert \Psi(t_{1}^{\prime})\right\rangle $ given by Eq. (\ref{bigpsi1}).
The pointers' positions in this case can be also be inferred from the
Monte-Carlo simulations computed above: since the probability to obtain a set
of positions $\mathbb{X}_{k}$ in state $\left\vert \Psi(t_{1}^{\prime
})\right\rangle $ does not depend on the measurement basis of qubit B, we can
use the simulations computed for a target function $\left\vert \left\langle
X_{1},...,X_{N}\right\vert \left.  \Phi_{A}^{\pm\theta}\right\rangle
\right\vert ^{2}$ (where $\theta$ can now be any angle) weighted by the
probability of occurrence of $\left\vert \Phi_{A}^{\pm\theta}\right\rangle $.
An illustration of the average distribution corresponding to the state
(\ref{manysp}) is illustrated in Fig. \ref{fig-Rprim} for $N=200$ (corresponding
in $\mathcal{R}$ to the distribution shown in Fig. \ref{fig-N200}).

Note that the subplots for $\left\vert \Phi_{A}^{\theta}\right\rangle $ or for
$\left\vert \Phi_{A}^{-\theta}\right\rangle $ obtained in $\mathcal{R}$ are
still meaningful in $\mathcal{R}^{\prime}$, not as a characterization of an
intermediate state $\left\vert \Phi_{A}^{\theta}\right\rangle $ or
$\left\vert \Phi_{A}^{-\theta}\right\rangle $ which does not exist in
$\mathcal{R}^{\prime}$ but as conditional distributions compatible with a
future measurement of $\hat{\sigma}_{\theta}$ on qubit B.\ For instance if
Alice$^{\prime}$ finds an average $\xi_{k}=-0.2$ when measuring $\mathbb{X}%
_{k}$, she can conclude that if $\hat{\sigma}_{z}$ is chosen to be measured on
qubit B, the $+1$ outcome will be obtained with certainty. Dealing with a
finite number of coupled pointers involves elements of both cases examined
previously: the states $\left\vert \Phi^{\epsilon_{\pm}}\right\rangle $ of Eq
(\ref{bigpsi1}) are not quasi-orthogonal, so that in $\mathcal{R}^{\prime}$
qubit B's state is updated to a state $\left\vert \psi_{B}^{\mathbb{X}_{k}%
}\right\rangle $ depending on the measured positions $\mathbb{X}_{k}$,
similarly to the state $\left\vert \psi_{B}^{X_{m}}\right\rangle $ of Eq.
(\ref{update-1wm}) after a single weak pointer is measured. There are however
a substantial number of sets $\mathbb{X}_{k}$ for which either $\left\langle
\mathbb{X}_{k}\right\vert \left.  \Phi^{\epsilon_{+}}\right\rangle $ or
$\left\langle \mathbb{X}_{k}\right\vert \left.  \Phi^{\epsilon_{-}%
}\right\rangle $ vanish, leading to a behavior similar to the
quasi-orthogonality or projective measurements. There are also sets for which
$\left\langle \mathbb{X}_{k}\right\vert \left.  \Phi_{A}^{+\theta
}\right\rangle $ or $\left\langle \mathbb{X}_{k}\right\vert \left.  \Phi
_{A}^{-\theta}\right\rangle $ can be nearly zero by interference [see Eqs.
(\ref{fipt})-(\ref{fimt})].

\section{Discussion and conclusion\label{disc}}

According to Peres, when measuring two qubits in an entangled state such as
$\left\vert \psi\right\rangle $ of Eq. (\ref{st1}) there are only two events,
the intervention on qubit A, and the intervention on qubit B while\ state
update ``is \emph{not} a physical process''
 \cite{peres-intervention}. Nevertheless, if the measured observable and the
outcome of qubit B are known (e.g., by classical communication) it
can be verified, without altering in any way the local qubit that the updated
state will be obtained with certainty. Hence it comes as no surprise that
views opposite to the one upheld by Peres can also be defended.\ For example
Ghirardi has pointed out \cite{ghirardi} that if one accepts that quantum
states are frame dependent (more precisely, they depend on hypersurfaces of
simultaneity) then it is consistent to uphold that state reduction is an
``objective'' process endowed with
Lorentz-covariant rules. Cohen and Hiley have instead argued \cite{hiley} that
assuming an objective instantaneous state
update model ruins any chance for a causal interpretation in terms of beables
depending on the quantum state properties. They show that introducing a
preferred frame is still consistent with quantum probabilities, except that in
this case the Hilbert space description at intermediate times breaks down in
frames other than the preferred frame, and the entire weak measurement process (including the unitary coupling) can only be described in the preferred frame.

At any rate, no-signaling appears as the practical impossibility for an
observer to extract any information concerning a remote measurement from the
updated state of the local qubit.\ While no-signaling is required for
consistency between descriptions in different reference frames, understanding
how no-signaling is enforced is tantamount to addressing the origin of
relativistic causality. Outcome randomness is most often cited as the main
ingredient in maintaining relativistic causality -- it is well-known that
no-signaling implies that quantum correlations must have random outcomes
\cite{PR,general} -- but we have seen when attempting to characterize
intermediate states with non-destructive weak measurements that superposition
comes into play.

In Sec. \ref{weakm} we have seen that the quantum state of a single weakly
coupled pointer is of the form
\begin{equation}
a\big\vert\varphi^{\epsilon_{+}}\big\rangle+b\big\vert\varphi^{\epsilon_{-}%
}\big\rangle \label{ppf}%
\end{equation}
where $a$ and $b$ depend on the distant (qubit B) measurement observable and
outcome and characterize the updated state. However since states
$\big\vert\varphi^{\epsilon}\big\rangle$ and $\big\vert\varphi^{\epsilon
^{\prime}}\big\rangle$ (where $\epsilon\neq\epsilon^{\prime}$) nearly overlap
[Eq. (\ref{overlap})] the superposition (\ref{ppf}) cannot be characterized in
a single shot, while repeating the procedure with many copies yields different
superpositions (at least stochastic outcomes for the same measured observable)
precluding any information to be extracted from the statistics of repeated
weak measurements. However if the distant qubit's measurement outcome is known, each
weak measurement can be sorted according to the updated state, and the
corresponding distribution (\ref{ppf}) can be obtained.

In order to circumvent the stochastic character of state update when repeating
measurements, we have introduced $N$ simultaneous weak measurements on a
multi-partite entangled state. When $N$ is very large however, we have shown
in Sec. \ref{inf} that the individually overlapping pointers behave
collectively as quasi-orthogonal pointers.\ Hence a superposition $a\left\vert
\Phi^{\epsilon_{+}}\right\rangle +b\left\vert \Phi^{\epsilon-}\right\rangle $
cannot be characterized in a single shot and the weak measurement becomes
equivalent to a projective measurement.

Finally, there is the intermediate case, in which there are sufficient
pointers to gather statistics but the collective pointer states are not
orthogonal and their superpositions overlap. The numerical simulations
presented in Sec. \ref{sev} indicate that quantitative partial information can
be obtained on the measurement outcomes made on the distant qubit. This
information however does not characterize the updated intermediate state but
the correlations of the entangled quantum state. For instance a
statistical quantity such as the average measured position $\xi$ obtained by
measuring in a single shot $N$ pointers correlates with a given probability
with an outcome on qubit B for any measured observable.\  This is a
conditional probability (that for some observables and a given realization of
$\xi$ may be zero or unity) correlating measurement outcomes observed by
distant parties, and does not imply that such a measurement has taken place
(say in $\mathcal{R}$) or will take place (in another frame
$\mathcal{R^{\prime}}$). Note it has been argued \cite{hofmann} that by
construction weak measurements decompose quite generally 
possible measurements into subensembles that are coherently reshuffled upon realization
of all the final measurements.

To conclude, we have seen that weak measurements -- like any other type of measurements -- characterize the correlations encapsulated in the full entangled state. An interplay between randomness and superposition (associated respectively with particle and wave characters) prevents any information from being acquired on the elusive intermediate frame-dependent state. No-signaling, implemented here through outcome randomness and superpositions appears as a fundamental property in maintaining consistency between descriptions in different reference frames.

\appendix
\counterwithin*{equation}{section}
\counterwithin*{figure}{section}

\setcounter{equation}{0}
\renewcommand{\theequation}{A\arabic{equation}}

\setcounter{figure}{0}
\renewcommand{\thefigure}{A\arabic{figure}}

\section*{Appendix: Metropolis-Hastings Sampling of a Multimodal Distribution}

\subsection*{Metropolis-Hastings Algorithm}

In Sec.\ \ref{sev}, we employ the Metropolis-Hastings (MH) algorithm \cite{monte}, a method from the Markov Chain Monte Carlo (MCMC) family, to sample from a complex, multivariable probability distribution. Direct sampling was not feasible due to the multimodal nature of the target distribution, which involved multiple peaks.

The implementation of the MH algorithm involves sampling a list of $N$ variables (points) from the multivariable distribution as follows:

\begin{enumerate}
	\item \textbf{Initialization}: We start by selecting an initial list of points $\mathbb{X}_{0} = (X_{0,1}, X_{0,2}, \ldots, X_{0,N})$ in the sample space, where each $X_{0,i}$ represents the starting value for variable $i$ in the chain.
	
	\item \textbf{Proposal Step}: At each iteration, a candidate list of points $\mathbb{X}^{*} = (X^{*}_{1}, X^{*}_{2}, \ldots, X^{*}_{N})$ is generated from a straightforward Gaussian proposal distribution $q(\mathbb{X}^{*}|\mathbb{X}_{t})$, where $\mathbb{X}_{t} = (X_{t,1}, X_{t,2}, \ldots, X_{t,N})$ is the current state of the chain. Each point in the list was proposed independently for each variable using a proposal distribution (identical for all variables) 
	\begin{equation}
	q_i(X_i^*|X_{t,i})  = \frac{1}{\sqrt{2\pi \sigma_q^2}} \exp\left( -\frac{(X_i^{*} - X_{t,i})^2}{2\sigma_q^2} \right)
	\end{equation}
	where $\sigma_q$ represents the standard deviation of the proposal distribution. 
	
	The overall proposal distribution is then given by 
	
	\begin{equation}
	q(\mathbb{X}^{*} | \mathbb{X}_{t}) = \prod_{i=1}^{N} q(X_i^{*} | X_{t,i}) = \prod_{i=1}^{N} \frac{1}{\sqrt{2\pi \sigma_q^2}} \exp\left( -\frac{(X_i^{*} - X_{t,i})^2}{2\sigma_q^2} \right)
	\end{equation}
	\item \textbf{Acceptance Step}: For the entire list of proposed points $\mathbb{X}^{*}$, we computed the acceptance ratio:
	\[
	\alpha = \min\left( 1, \frac{\pi(\mathbb{X}^{*}) q(\mathbb{X}_{t}|\mathbb{X}^{*})}{\pi(\mathbb{X}_{t}) q(\mathbb{X}^{*}|\mathbb{X}_{t})} \right)
	\]
	where $\pi(\mathbb{X})$ represents the target multivariable distribution, and $q$ is the proposal distribution. 
	
	Since $q$ is a symmetric function, we get $q(\mathbb{X}_{t}|\mathbb{X}^{*})=q(\mathbb{X}^{*}|\mathbb{X}_{t})$, and thus, the acceptance ration becomes 
	
	\begin{equation}
	\alpha = \min\left( 1, \frac{\pi(\mathbb{X}^{*})}{\pi(\mathbb{X}_{t}) } \right)
	\end{equation}
	This acceptance ratio determines whether the entire new list of variables should be accepted.
	
	\item \textbf{Move or Stay}: With probability $\alpha$, we either accept the candidate list $\mathbb{X}^{*}$ (setting $\mathbb{X}_{t+1} = \mathbb{X}^{*}$) or retain the current list $\mathbb{X}_{t+1} = \mathbb{X}_{t}$. This can be simply achieved by generating a random variable $y \in [0,1]$, and accepting $\mathbb{X^*}$ if \  $y< \alpha$.
	
	\item \textbf{Iterate}: We repeat these steps for several iterations. In each iteration, a new sample list of size $N$ is generated (equal to the number of variables) and the average of the list across iterations is recorded.
\end{enumerate}

\subsection*{Wave Function and Target Distribution}

\begin{figure}[H]
\centering
\includegraphics[width=0.6\textwidth]{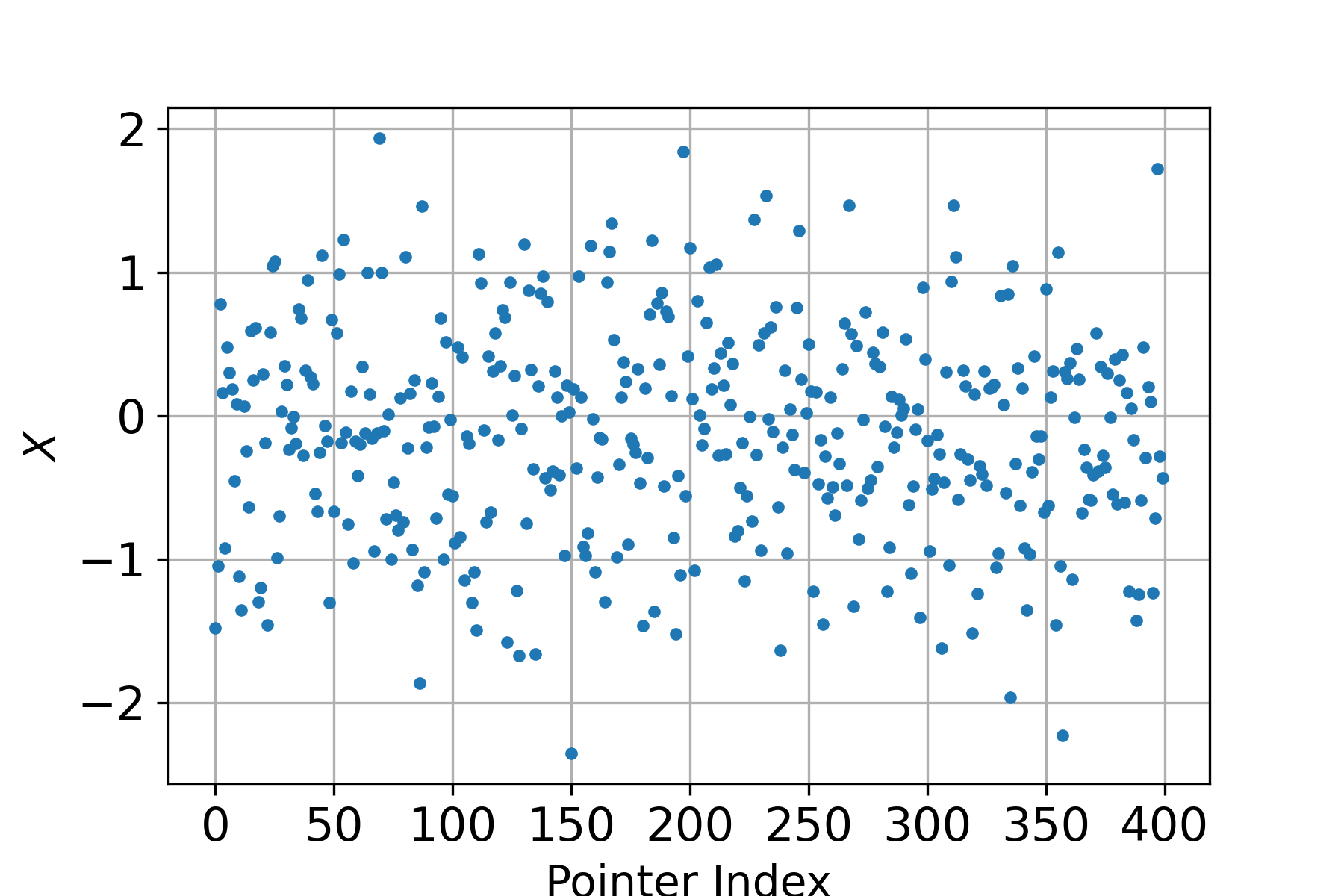}\caption{A sample from the distribution for the updated state given by Eq.\ (\ref{fimt}) (representing the case in which Bob measures in the $\pm \theta$ basis and obtains $-\theta$, corresponding to Fig. \ref{400} (b)). The plot gives the measured positions of each pointer (here $N=400$). In this specific case, the average of the sample is  $\xi=-0.12$.}%
\label{sample-fig}%
\end{figure}

The target distribution is derived from the absolute square of a wave function of $N$ identical pointers (as in Eq.(\ref{superif})), expressed as:

\begin{equation}
	\Phi(\mathbb{X}) = \left\langle \mathbb{X} | \Phi\right\rangle = a \prod_{i=1}^{N} \exp\left( -\frac{(X_{i}-\epsilon)^{2}}{2d^{2}} \right) + b \prod_{i=1}^{N} \exp\left( -\frac{(X_{i}+\epsilon)^{2}}{2d^{2}} \right)
\end{equation}

where $a$ and $b$ are normalization constants, $\epsilon$ represents a shift, and $d$ is the standard deviation of the Gaussians. Remember that the individual pointer states are given by 
\begin{equation}
\varphi_i(X_i\pm\epsilon)=\frac{1}{\sqrt{2\pi d^2}}\exp\bigg(-\frac{(X_i\pm \epsilon)^2}{2d^2}\bigg)
\end{equation} 
Note that we set $d=1$, and the shift in the weak pointer's position was chosen to be $\epsilon=g=0.1$.  The resulting target distribution was:

\begin{equation}
	\pi (\mathbb{X})=|\Phi(\mathbb{X})|^{2} = \left( a \prod_{i=1}^{N} \exp\left( -\frac{(X_{i}-\epsilon)^{2}}{2} \right) + b \prod_{i=1}^{N} \exp\left( -\frac{(X_{i}+\epsilon)^{2}}{2} \right) \right) ^{2}
\end{equation}
where $a$ and $b$ depend on the given updated state (see Eqs.\ (\ref{fipw})-(\ref{fimt}) and (\ref{manysp})). A specific sample is illustrated in Fig.\ \ref{sample-fig}.

\subsection*{Considerations for Proposal Distribution}

Even though our target distribution is multimodal, with two distinct peaks, this straightforward approach turned out to be sufficient for sampling from the multivariable distribution, given the problem constraints. A key factor for successful sampling is choosing the width of the proposal distribution $\sigma_q$, as one needs to tune the width of the ``jumping" steps to guarantee that the proposed point lies within a reasonably probable region. The choice of $\sigma_q$ becomes more critical as the number of variables (weak pointers) increases, since the peaks of the distribution of $\xi$ become narrower, making it harder to achieve a successful jump (compare Figs. \ref{fig-N200} and \ref{400}).

\subsection*{Results and Discussion}

 The histograms of the sample means ($\xi$) were scaled to represent the total probability of obtaining outcomes associated with the states studied (depending on Bob's choice of measurement and outcome).

One of the primary challenges was ensuring accurate sampling across the multimodal distribution. As the number of variables (or weak pointers) increases, the likelihood of successfully sampling from both peaks, or jumping between them, decreases. To address this, we used a large number of samples to ensure the Monte Carlo process stabilized within the target multivariable distribution.


\begin{thebibliography}{99}                                                                                               %


\bibitem {gisin-science}N.\ Gisin, Science 326 1357 (2009).

\bibitem {BT}J. S. Bell, \ Physics Physique Fizika, \ 1,\ 195 (1964)

\bibitem {Albert-Aharonov-1}Y. Aharonov and D. Z. Albert, \ Phys. Rev. D \ 24,
359 (1981).
\bibitem {peres-intervention}A. Peres, \ Phys. Rev. A \ 61 022117 (2000)

\bibitem {shimony}A. Shimony, \ International Philosophical Quarterly 18(1),
\ 3 (1978)



\bibitem {bell-ns}J. S Bell, \emph{La Nouvelle Cuisine}, in Between Science
and Technology, editors A. Sarlemijn and P. Kroes. Elsevier Science Publishers (1990).


\bibitem {AA2}Y. Aharonov and D. Z. Albert, Phys. Rev. D 29, 228 (1984).

\bibitem {Popescu-Vaidman}S. Popescu and L. Vaidman, Phys. Rev. A 49, 4331 (1994).

\bibitem {Reznik}B. Groisman and B. Reznik, Phys. Rev. A 66, 022110 (2002).

\bibitem {brodutch}A. Brodutch and E. Cohen, Phys. Rev. Lett. 116, 070404 (2016).

\bibitem {exp-time}F. Garrisi et al., Sci Rep 9, 11897 (2019).

\bibitem {srikanth}R.\ Srikanth, Phys.\ Rev.\ A 106, 012221 (2022).


\bibitem {javurek}D.\ Javurek, Phys. Rev. A 108, 023709 (2023)


\bibitem {AMWF}J. Allam and A. Matzkin EPL 143 60001 (2023).

\bibitem {peres-review}A. Peres and D. R. Terno, \ Rev. Mod. Phys. 76, \ 93 (2004)

\bibitem {HK}K. E. Hellwig, \ and K. \ Kraus, \ Phys. Rev. D 1, 566 (1970)


\bibitem {hiley}O. Cohen and B. J. Hiley,\ Found. Phys. 25, \ 1669 (1995)


\bibitem {PR}S. Popescu and D. Rohrlich, Found. Phys. 24, 379 (1994) .

\bibitem {general}L. Masanes, A. Acin and N. Gisin, Phys.\ Rev.\ A 73, 012112 (2006).


\bibitem {AAV}Y. Aharonov and D. Z. Albert and L. Vaidman, \ Phys. Rev. Lett.
60, \ 1351 (1988).

\bibitem{bey}B. E. Y. Svensson, Found Phys  45, 1645 (2015)

\bibitem{vaidman-con} L. Vaidman, Phil. Trans. R. Soc. A.375, 20160395 (2017)

\bibitem {FP} A. Matzkin, Found Phys 49, 298 (2019).


\bibitem{JJJ} J. R Hance, M. Ji and H. F. Hofmann, New J. Phys. 25 113028 (2023) 

\bibitem {lundeen} J. S. Lundeen, B. Sutherland, A. Patel, C. Stewart and C. Bamber, Nature 474 188 (2011).

\bibitem {PRL127}L. Xu et al., Phys. Rev. Lett. 127, \ 180401 (2021)

\bibitem {ion} Y. Pan et al., Nat. Phys. 16, 1206 (2020).

\bibitem {unambiguous} S. N. Sahoo et al.,  Commun. \ Phys. \ 6, 203 (2023)

\bibitem{DMMJB} J. Dressel et al.,  Rev. \ Mod. \ Phys. 86, 307 (2014)

\bibitem{TC} B. Tamir and E. Cohen, Quanta 2: 7-17 (2013)

\bibitem {sudarshan} I. M. Duck, P. M. Stevenson, and E. C. G. Sudarshan, Phys. Rev. D 40, 2112 (1989).

\bibitem {tollaks} J. Tollaksen  J. Phys.: Conf. Ser. 70 012015 (2007).

\bibitem {ahar-reznik}Y. Aharonov and B. Reznik, \ Phys. Rev. A 65, \ 052116 (2002)

\bibitem {finkelstein}J. Finkelstein, \ Phys. Rev. A 67, \ 026101 (2003)

\bibitem {ghirardi}G. C. Ghirardi, Found. Phys. 30, 1337 (2000).

\bibitem{hofmann} H. H. Hofmann, Phys. Rev. A 81, 012103 (2010).

\bibitem{monte} J. E. Gentle, Statistics and computing - Random Number Generation and Monte Carlo Methods [2nd ed] (p. 140), Springer (2003)







\end{thebibliography}
\end{document}